\documentclass[twocolumn,aps,pra,showpacs,floatfix]{revtex4}

\usepackage{graphicx}
\usepackage{amsmath}
\usepackage{bm}
\usepackage[]{natbib}

\begin{document}

\bibliographystyle{myprsty}

\title{Phase-dependent interference mechanisms in a three-level 
$\Lambda$ system driven by a quantized laser field}

\author{J\"org Evers}
\email{joerg.evers@mpi-hd.mpg.de}
\affiliation{Max-Planck-Institut f\"ur Kerphysik, Saupfercheckweg 1, D-69117
Heidelberg, Germany}

\begin{abstract}
The dynamics of an atomic few-level
system can depend on the phase of driving fields coupled to the 
atom if certain conditions are satisfied. This is of particular 
interest to  control interference  effects, which can alter
the system properties considerably. In this article, we discuss
the mechanisms of such phase control and interference effects in
an atomic three-level system in $\Lambda$ configuration,
where the upper state spontaneously decays into the two lower states.
The lower states are coupled by a driving field, which we treat 
as quantized. This allows for an interpretation on the single photon 
level for both the vacuum and the driving field.
By analyzing the system behavior for a driving field initially in 
non-classical states with only few Fock number states populated, we find 
that even though the driving field is coupled to the lower states only, it 
induces a multiplet of upper states. Then interference occurs independently
in three-level subsystems in $V$ configuration, each formed by two 
adjacent upper states and a single dressed lower state. 
\end{abstract}

\pacs{42.50.Lc,42.50.Ct,42.50.Hz}

\maketitle

\section{Introduction}
Interference is recognized as a major mechanism to control
quantum dynamics~\cite{FiSw2005}. 
If a system exhibiting quantum interference is sensitive 
to the phase of applied driving fields, then the interference itself 
can often be conveniently controlled.
This is intuitively clear, as the relative phase between
the interfering pathways decides whether constructive or destructive
interference takes place. Thus it is not surprising that phase-dependent
systems have been discussed
extensively~\cite{MaHeSaNaKe1997,PaKn1998,GhZhZu2000,GaLiZh2002,
WuGa2002,QuWoJoAg1997,GoPaKn1998,FiSeSoAd2000,MaSo2004,MoFrOp2002,XuWuGa2004,
KoLeHuBaWi2000,BoWiFr2001,
MeAg1998,MaEvKe,MaEv2004,FiSw2005,SaTaKiZu2004}. The phase-control
of quantum interference has also been verified 
experimentally~\cite{KoLeHuBaWi2000}.
One of the first and simplest examples for a phase-sensitive
system is a three-level system in $\Lambda$ configuration, where
the upper state decays spontaneously into the two lower
states coupled by a classical driving field~\cite{MaHeSaNaKe1997}.
The transition dipole moments from the upper state to the two
lower states are assumed to be non-orthogonal.
Then, due to the driving field, 
each of the two lower states can be reached via two
pathways: Either by a direct spontaneous decay, or by a decay into
the other state followed by a driving field-induced transition.
The two paths interfere, with a relative phase between the two
path amplitudes equal to the driving field phase.
This classical interpretation, however, is unsatisfactory for
several reasons.
First, the respective initial and the final states of two interfering pathways
are obviously not the same, as one involves an interaction with
the driving field, whereas the other does not. 
Thus one could
argue that the two paths do not interfere, as they could be 
distinguished by a measurement. A typical counter-argument against this
is that the photon number distribution of a strong coherent driving 
field has a large width, such that the two paths cannot be distinguished.
This argument, however, cannot be verified using a semi-classical
description of the system.
Second, while it is clear that the two pathways must have
either a different initial or a different final state, it is not
apparent in the semiclassical description what the exact pathways are. 
Third, the classical description restricts the system to
classical driving fields, and thus does not allow to consider
non-classical driving fields to further study the system properties. 
Finally, it is not obvious where the phase of the driving field should
be considered in a quantized treatment. For example, in the quantized
treatment, both the coupling
constants and the initial field state can carry phase information.

\begin{figure}[b!]
\includegraphics[width=5.5cm]{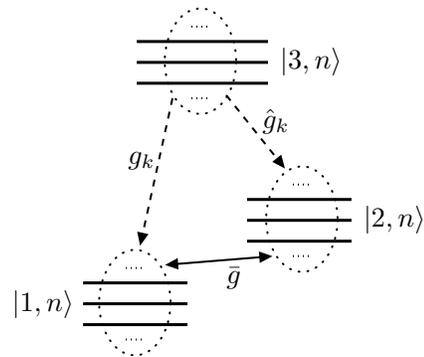}
\caption{\label{fig-system}Bare-state representation
of the considered system. The multiple lines for each state
denote the Fock state multiplets induced by the coupling of the
atom with a quantized driving field. The upper states $|3,n\rangle$ decay
spontaneously into the two lower state multiplets $|1,n\rangle$ and
$|2,n\rangle$. Here, $n$ is the number of photons in the respective
driving field Fock modes.
The two lower state multiplets are coupled by the quantized
driving field. $g_k$ and $\hat{g}_k$
are coupling constants for the interaction with the vacuum giving
rise to spontaneous emission, $\bar{g}$
is the coupling constant for the driving field.}
\end{figure}

Thus in this article, we revisit the three-level system  
in $\Lambda$ configuration, but treat the driving field as
quantized. This allows to discuss the phase and 
interference effects on the single photon level.
We derive a set of equations for the two dressed lower states alone, 
and solve this numerically for different field configurations. 
For coherent fields, which are closest to classical fields, we find 
the same results as in the classical case.
By analyzing driving fields consisting of a single Fock mode, 
few adjacent number states, or separated Fock states, it is possible
to find the origin of the quantum interference and to identify the
interfering pathways. It turns out that it is not the large total 
width of the photon number distribution which is crucial for the 
interference, but 
rather a fixed phase relation (coherence) between adjacent Fock modes. 
The driving field gives rise to an upper
state multiplet, even though it is only coupled to the lower 
state~\cite{EvKe2002b,EvKe2004a,AkEvKe2005}.
Then the interference occurs independently in three-level 
subsystems in $V$-configuration, which are formed by two adjacent upper 
states and a single lower dressed state. 
Finally, some insight is gained in the role of phase in quantum
systems by comparing the quantum treatment with the semiclassical
description.

The article is organized as follows: In the next 
section~\ref{sec-analytic}, we analytically 
derive a set of equations for the two dressed ground states alone,
which is then solved numerically in Section~\ref{sec-numerics}
for different initial field configurations. The results
of this analysis are discussed and summarized in 
Section~\ref{sec-discussion}.

\section{\label{sec-analytic}Analytical considerations}
In the following, we consider a three-level system in 
$\Lambda$-configuration as shown in Fig.~\ref{fig-system}.
The upper state $|3\rangle$ decays to the two
lower states $|1\rangle$, $|2\rangle$ with decay rates
$\gamma_1$, $\gamma_2$, respectively. In addition, 
the two lower states are connected by a coherent field,
which in the following we treat as a quantized field.
For simplicity, we assume the driving field to be on 
resonance with the transition frequency $\omega_{21}$ between
the two lower states.
The interaction Hamiltonian in the interaction picture may then be 
written as~\cite{scullybook}
\begin{align}
V(t) =& V_{\rm drive} + V_{\rm SE}(t)\,,
\end{align}
where
\begin{align}
V_{\rm drive} =& i \hbar \left ( \bar{g} A_{21} b 
    - \bar{g}^* A_{12} b^\dagger \right ) \,,\\
V_{\rm SE}(t) =& i \hbar \sum_k \left ( g_k A_{31} a_k e^{-i\delta_k t}- 
    g_k^* A_{13} a_k^\dagger e^{i\delta_k t} \right .\nonumber \\
    &\left . + \hat{g}_k A_{32} a_k e^{-i\hat{\delta}_k t}- 
    \hat{g}_k^* A_{23} a_k^\dagger e^{i\hat{\delta}_k t} \right )\,.
\end{align}
Here, $V_{\rm SE}(t)$ denotes the interaction of the atom with the 
surrounding vacuum field giving rise to spontaneous emission, 
and $V_{\rm drive}$ describes the interaction with the quantized
driving field. $A_{ij} = |i\rangle\langle j|$ ($i,j \in \{1,2,3\}$) are
atomic transition operators, $b$ ($b^\dagger$) are annihilation
(creation) operators of the driving field, and $a_k$ ($a_k^\dagger$)
are the corresponding operators of the surrounding vacuum mode $k$.
$\delta_k = \omega_k - \omega_{31}$ and $\hat{\delta}_k = 
\omega_k - \omega_{32}$ are detunings to atomic transition frequencies
$\omega_{ij} = \omega_i - \omega_j$, where $\hbar \omega_i$ is the
energy of state $|i\rangle$. $\bar{g}, g_k$ and $\hat{g}_k$ are
coupling constants. As we are interested in the phase-dependence
of the system, these constants are treated as complex entities. In 
particular, we define
\begin{align}
\bar{g} = |\bar{g}|\, e^{i\phi}\,.
\end{align}
Note that for simplicity, we treat the transitions
between the two lower atomic states as one-photon
 transitions~\cite{MaHeSaNaKe1997}.
However, replacing $V_{\rm drive}$ with a corresponding effective
Hamiltonian containing two-photon transitions proportional to
$b b$ ($b^\dagger b^\dagger$) does not change the involved physics.
The interference effects in this system crucially depend on the
relative orientation of the dipole moments of the transitions
$|3\rangle \leftrightarrow |1\rangle$ and  
$|3\rangle \leftrightarrow |2\rangle$.
If the two dipole moments are orthogonal, then no interference effects
are possible. A simple interpretation for this result is that then
the photons emitted spontaneously on the two transitions can be 
distinguished by their polarizations, such that no pathway interference
is possible. Thus in the following, we assume the two dipole moments
to be parallel. Note that spontaneously generated coherences between
the two lower states which arise if further 
$\omega_{21}$ is smaller than or of the same order as $\gamma_1, \gamma_2$ 
do not play a role in the current system setup. These coherences, 
however, become relevant
if additional driving fields on transitions $|3\rangle \leftrightarrow
 |i\rangle$ ($i\in\{1,2\}$) couple to them (see, e.g., \cite{EvBuKe2002}).
\begin{figure}[t]
\includegraphics[height=5cm]{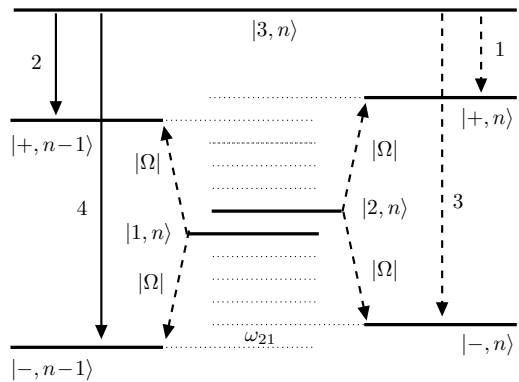}
\caption{\label{fig-system-dress}Dressed state representation
of the system configuration with a quantized driving field.
The two lower bare states $|1\rangle$, $|2\rangle$ are
AC-Stark shifted and absorbed into the dressed states
due to the driving field. The numbers at the arrows
correspond to the numbers at the peaks in Fig.~\ref{fig-noint}.}
\end{figure}

To simplify the analysis, we transfer the system to the dressed
state representation for the two driven states, which is given by
\begin{subequations}
\label{dressedstates}
\begin{align}
|+,n\rangle =& \frac{1}{\sqrt{2}}\left ( |1,n+1\rangle
+ i\,e^{i\phi}\, |2,n\rangle   \right )\,,\\
|-,n\rangle =& \frac{1}{\sqrt{2}}\left (|1,n+1\rangle
- i\,e^{i\phi}\, |2,n\rangle  \right )\,. 
\end{align}
\end{subequations}
The dressed-state representation of the system is shown in 
Fig.~\ref{fig-system-dress} for parameters where the
AC-Stark splitting is larger than the level splitting of
the two lower bare states.

The wave function of the system can be written as
\begin{align}
|\Psi(t)\rangle =& \sum_n C_3^n(t)\,|3,n,0\rangle
 + \sum_{n,k}  \alpha_k^n(t)\,a_k^\dagger\,|+,n,0\rangle 
 \nonumber \\
 &+ \sum_{n,k} \beta_k^n(t)\,a_k^\dagger\,|-,n,0\rangle\,.
\end{align}
Here, $|i,n,0\rangle$ denotes a state with no photons
in the surrounding vacuum, $n$ photons in the driving 
field mode and the atom in electronic state $i$ ($i=3,+,-$).
From the Schr\"odinger equation,
\begin{align}
i\hbar \frac{d}{dt}\, |\Psi(t)\rangle = V(t)\, |\Psi(t)\rangle\,,
\end{align}
the equations of motion for the state amplitudes can be evaluated
to give
\begin{align}
\frac{d}{dt}&\, C_3^n(t) = \frac{1}{\sqrt{2}} \sum_k g_k \,
e^{-i\delta_k t}\,  \left( \alpha_k^n(t) + \beta_k^n(t)\right )
\nonumber \\
 & + \frac{i}{\sqrt{2}} \sum_k \hat{g}_k \,
e^{i\phi}\,e^{-i\hat{\delta}_k t}\,  \left( \alpha_k^n(t) - \beta_k^n(t)\right )
 \,, \label{eom-c}\\
\frac{d}{dt}&\, \alpha_k^n(t) = - \frac{1}{\sqrt{2}} \, C_3^n(t) 
\left ( g_k^* \,e^{i\delta_k t} + i \hat{g}_k^*\, e^{-i\phi} \, 
e^{i\hat{\delta}_k t}
\right )\nonumber \\
& - \frac{i}{2} \, |\bar{g}|\, \sqrt{n+1} \left ( 
 \alpha_k^{n+1}(t) + \beta_k^{n+1}(t) \right )
 \nonumber \\
& - \frac{i}{2} \, |\bar{g}|\, \sqrt{n}\, \left ( 
 \alpha_k^{n-1}(t)  - \beta_k^{n-1}(t) 
  \right )\,,\label{eom-alpha}\\
\frac{d}{dt}&\, \beta_k^n(t) =  - \frac{1}{\sqrt{2}} \, C_3^n(t) 
\left ( g_k^* \,e^{i\delta_k t} -i \hat{g}_k^*\, e^{-i\phi} \, 
e^{i\hat{\delta}_k t}
\right )\nonumber \\
&  + \frac{i}{2} \, |\bar{g}|\, \sqrt{n+1} \left ( 
 \alpha_k^{n+1}(t) + \beta_k^{n+1}(t) \right )
 \nonumber \\
& - \frac{i}{2} \, |\bar{g}|\, \sqrt{n}\, \left ( 
 \alpha_k^{n-1}(t)  - \beta_k^{n-1}(t) 
  \right )\,.\label{eom-beta}
\end{align}
As shown in Appendix~\ref{app-derivation}, in a Wigner-Weisskopf-like
evaluation, and assuming that the atom initially is in the upper
state $|3\rangle$ ($\alpha_k^n(0)=0=\beta_k^n(0)$), 
one may solve the equation of 
motion for the upper state amplitudes to give
\begin{align}
C_3^n(t) = C_3^n(0)\,e^{-\frac{1}{2}\gamma t}\,.\label{sol-c3}
\end{align}
The rate $\gamma=\gamma_1+\gamma_2$ is the total decay rate of the 
upper state $|3\rangle$
to both lower states.
Note that each of the Fock number state amplitudes of the driving 
field decays independently. The reason is that the spontaneous emission
from the upper state is an irreversible process, such that
the driving field coupling to the lower states cannot influence
the decay of the upper states. As the Wigner-Weisskopf procedure involves
the continuum limit of an infinite quantization volume, in the following,
we replace the discrete ground state amplitudes and the detunings
by their continuous counterparts:
\begin{align}
\alpha_k^n(t) &\longrightarrow \alpha_\omega^n(t)\,, \qquad
\beta_k^n(t) \longrightarrow \beta_\omega^n(t)\,,\\
\delta_k &\longrightarrow \delta_\omega = \omega - \omega_{31}\,,\\
\hat{\delta}_k &\longrightarrow \hat{\delta}_\omega = \omega - \omega_{32}\,.
\end{align}
On inserting Eq.~(\ref{sol-c3}) in Eqs.~(\ref{eom-alpha}), (\ref{eom-beta}),
for each spontaneous emission frequency $k$, one obtains a set of 
coupled equations for the ground state coefficients $\alpha_k^n(t)$ 
and $\beta_k^n(t)$ alone, which can be solved
numerically for any given initial state:
\begin{align}
\frac{d}{dt}&\, \alpha_\omega^n(t) = - \frac{1}{\sqrt{2}} \, 
C_3^n(0) \, e^{-\frac{1}{2}\gamma t}\,
\left \{ g(\omega_{31})^* \,e^{i\delta_\omega t} \right .\nonumber \\
&\left . + i \hat{g}(\omega_{32})^*\, e^{-i\phi} \, 
e^{i\hat{\delta}_\omega t}
\right \}\nonumber \\
& - \frac{i}{2} \, |\bar{g}|\, \sqrt{n+1} \left ( 
 \alpha_\omega^{n+1}(t) + \beta_\omega^{n+1}(t) \right )
 \nonumber \\
& - \frac{i}{2} \, |\bar{g}|\, \sqrt{n}\, \left ( 
 \alpha_\omega^{n-1}(t)  - \beta_\omega^{n-1}(t) 
  \right )\,,\label{sys-alpha}\\
\frac{d}{dt}&\, \beta_\omega^n(t) =  - \frac{1}{\sqrt{2}}\, 
C_3^n(0) \, e^{-\frac{1}{2}\gamma t}\, 
\left \{ g(\omega_{31})^* \,e^{i\delta_\omega t} \right. \nonumber \\
&\left. -i \hat{g}(\omega_{32})^*\, e^{-i\phi} \, 
e^{i\hat{\delta}_\omega t}
\right \}\nonumber \\
&  + \frac{i}{2} \, |\bar{g}|\, \sqrt{n+1} \left ( 
 \alpha_\omega^{n+1}(t) + \beta_\omega^{n+1}(t) \right )
 \nonumber \\
& - \frac{i}{2} \, |\bar{g}|\, \sqrt{n}\, \left ( 
 \alpha_\omega^{n-1}(t)  - \beta_\omega^{n-1}(t) 
  \right )\,.\label{sys-beta}
\end{align}
Note that for typical atomic setups, one has 
$g(\omega_{31}) \approx \hat{g}(\omega_{32})$, which we assume in 
the following.
From the ground state amplitudes, the spontaneous emission spectrum
$S(\omega)$ may then be evaluated as
\begin{align}
S(\omega) \propto \sum_n \left ( |\alpha_\omega^n(t\to \infty)|^2 
   + |\beta_\omega^n(t\to \infty)|^2 \right )\,, \label{spectrum}
\end{align}
where $\alpha_\omega^n(t\to \infty)$ and 
$\beta_\omega^n(t\to \infty)$ are the steady states of the ground state
amplitudes.
It is interesting to note that the expression for the emission spectrum
Eq.~(\ref{spectrum}) includes contributions both from the independent
decay of the two transitions from the upper state and of interference
terms, though somewhat hidden. On expanding Eq.~(\ref{spectrum}),
one obtains terms proportional to $|g(\omega_{31})|^2$ and to 
$|\hat{g}(\omega_{32})|^2$, which correspond to the independent 
spontaneous decay rates $\gamma_1$ and $\gamma_2$, respectively.
Additionally, the spectrum contains contributions proportional to
$g(\omega_{31})\hat{g}(\omega_{32})^*$ and 
$g(\omega_{31})^*\hat{g}(\omega_{32})$, which are the interference 
terms. If the two transition dipole moments $|3\rangle \leftrightarrow
|i\rangle$ ($i\in\{1,2\}$) are assumed to be orthogonal, then these 
interference contributions vanish.

\section{\label{sec-numerics}Numerical analysis}
\begin{figure}[t]
\includegraphics[width=8cm]{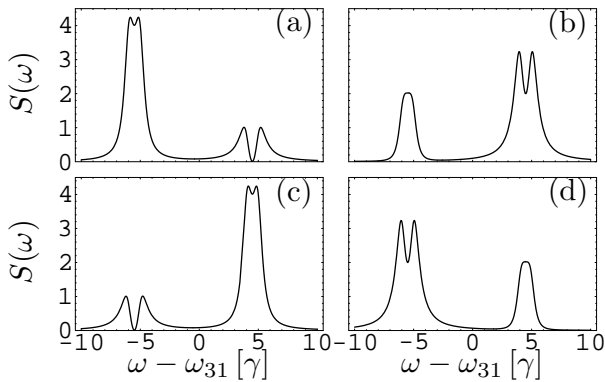}
\caption{\label{fig-reference}Reference figures obtained using a classical
driving field. $S(\omega)$ is the spontaneous emission spectrum of the system
shown in Fig.~\ref{fig-system}. The parameters are 
$|\Omega| = 5\,\gamma, \omega_{21}=\gamma$,
and the driving field phase $\phi_c$ is (a) $\phi_c = 0$, (b)  
$\phi_c = 0.5 \,\pi$
and (c) $\phi_c = \pi$ and (d) $\phi_c = 1.5\,\pi$.}
\end{figure}

In this section, we numerically solve the set of equations
Eqs.~(\ref{sys-alpha}),(\ref{sys-beta}) for several cases
of the initial distribution of Fock number states of the driving
field. In particular, we compare the results to those obtained
from a calculation involving a classical driving field,
where $V_{\rm drive}$ is replaced by the classical interaction picture 
Hamiltonian~\cite{MaHeSaNaKe1997}
\begin{align}
V_{\rm drive}^{\rm class.} = i\hbar \left ( \Omega \, A_{21} 
- \Omega^* \, A_{12}\right ) \,. \label{classic}
\end{align}
Here, $\Omega = |\Omega|\,e^{i\phi_c}$ is the  Rabi
frequency for a classical field with phase $\phi_c$.
Using the classical Hamiltonian Eq.~(\ref{classic}), for
$|\Omega| = 5\,\gamma$ and $\omega_{21}=\gamma$, we obtain
four reference figures shown in Fig.~\ref{fig-reference}
for (a) $\phi_c = 0$, (b)  $\phi_c = 0.5 \,\pi$
and (c) $\phi_c = \pi$ and (d) $\phi_c = 1.5\,\pi$. (c.f.
Fig. 4 in \cite{MaHeSaNaKe1997}).

\subsection{Coherent field}
We start the numerical analysis of the system driven by a quantized field
by assuming that initially the driving field is in a coherent state
\begin{align}
C_3^n(0) = e^{-|\alpha|^2/2} \frac{\alpha^n}{\sqrt{n!}}\,, \qquad 
\alpha = |\alpha|\,e^{i\phi_\alpha}\,.
\end{align}
We thus have two phases in our system: $\phi$ is the phase of the 
coupling constant
$\bar{g}$, and $\phi_\alpha$ is the phase of the initial coherent state.
Note that $|\alpha|^2$ is the average number of photons in the driving 
field mode,
which for strong coherent fields can be used to relate the classical Rabi
frequency  to the quantum counterpart via the equation 
$|\Omega| = |\bar{g}|\cdot|\alpha|$, as $|\alpha| \approx |\alpha|\pm 1$ 
for $|\alpha|\gg 1$ in the semiclassical limit.
Then, using the same values for $|\Omega|$ and $\omega_{21}$ as in the 
reference 
Fig.~\ref{fig-reference}, we vary the phases $\phi$ and $\phi_\alpha$. 
The results are shown in Table~\ref{tab-1}. In summary, the quantum 
simulation with a coherent state as initial state of the driving field 
mode yields identical results to the classical calculation if one identifies 
the phases as follows:
\begin{align}
\phi + \phi_\alpha \leftrightarrow \phi_c\,. \label{phase-rel}
\end{align}

One may intuitively understand this result by
taking the expectation value of the
quantum interaction Hamiltonian $V_{\rm drive}$ for the coherent 
state $|\alpha\rangle$:
\begin{align}
\langle \alpha | \, V_{\rm drive} \, |\alpha \rangle = 
i\hbar \left (\bar{g}\,\alpha\,A_{21} - \bar{g}^*\,\alpha^*\,A_{12}\right)
\end{align}
A comparison with the classical Hamiltonian Eq.~(\ref{classic}) again yields
$|\Omega| = |\bar{g}|\cdot|\alpha|$ and the phase 
relation Eq.~(\ref{phase-rel}).
Thus, in a quantized description of a classical driving field with a 
fixed phase, the classical phase may either be attributed to the coupling 
constants or to the initial field state.

\begin{table}[t]
\begin{ruledtabular}
\begin{tabular}{|c|c|c|c|}
\hline 
& $\phi = 0$ & $\phi = \pi/2$ & $\phi = \pi$ \\  
\hline
\hline
$\phi_\alpha = 0$     & a) & b) & c) \\
$\phi_\alpha = \pi/2$ & b) & c) & d) \\
$\phi_\alpha = \pi$   & c) & d) & a) \\
\hline 
\end{tabular}
\caption{\label{tab-1}Results of the numerical simulation
of the system in Fig.~\ref{fig-system} driven by a quantized field
which initially is in a coherent state with phase $\phi_\alpha$.
$\phi$ is the phase of the coupling constant $\bar{g}$. In the table,
e.g. a) means that the result for this phase configuration 
is identical to reference a) in Fig.~\ref{fig-reference}.}
\end{ruledtabular}
\end{table}

\subsection{Single Fock mode}
\begin{figure}[b]
\includegraphics[width=6cm]{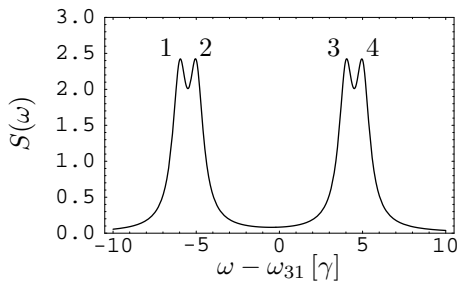}
\caption{\label{fig-noint}Spontaneous emission spectrum without interference.
This figure is either obtained for a single Fock mode as initial
state of the driving field, or by averaging the spectrum for
a coherent driving field over all possible values of the phase.
The parameters are as in Fig.~\ref{fig-reference}. The numbers at the
peaks correspond to the transitions indicated in the dressed state
picture of the system in Fig.~\ref{fig-system-dress}.}
\end{figure}

We now study the system where the quantized driving field initially is in 
a single Fock mode:
\begin{align}
C_3^n(0) = \delta_{n,n_0}\,.
\end{align}
Here $\delta_{ij}$ is the Kronecker delta symbol, and $n_0$ is 
the photon number of the Fock state initially populated.
The resulting spontaneous emission spectrum is shown in 
Fig.~\ref{fig-noint}. It does not depend on the phase
$\phi$, which is reasonable, as a number state does not
possess a definite phase~\cite{Lo2000}.
The obtained spectrum is essentially a sum of four independent
Lorentzian structures which correspond to the four transitions
from the upper state to the corresponding  dressed states as
shown in Fig.~\ref{fig-system-dress}.  
It turns out that the spectrum for a single Fock state
is identical to the spectrum obtained for a coherent or classical
driving field if the interference effects are ignored.
This can either be done by removing the corresponding terms
from the equations of motion, or by averaging the emission
spectrum over all possible values of the phase $\phi$.
Thus no interference occurs for a single Fock mode.

\subsection{Adjacent Fock modes}
Next we consider a driving field where few adjacent 
Fock states are initially populated with a fixed phase
relation, for simplicity with equal weight:
\begin{align}
C_3^n(0) = \frac{1}{\sqrt{2W+1}}\,\sum_{k=-W}^{W}\,
e^{i\,k\,\phi_\alpha t}\, \delta_{n,n_0+k}\,.
\end{align}
Here, $W$ is the width of the initial Fock mode distribution
around photon number $n_0$.
For this initial configuration, the resulting emission
spectra are qualitatively similar to the ones shown in 
Fig.~\ref{fig-reference}, even for $W=1$. In particular,
the phase dependence is the same. Quantitatively,
there are deviations from the classical spectrum for low 
$W$, which disappear with increasing $W$. The reason for
these deviations is given in Sec.~\ref{sec-discussion}.

\subsection{Separated Fock modes}
Finally we study the case where $N$ non-adjacent Fock states 
are initially populated, again with equal weight for simplicity:
\begin{align}
C_3^n(0) = \frac{1}{\sqrt{N}}\,\sum_{k=1}^{N}\,e^{i\,\kappa_k\,\phi_\alpha t}\, \delta_{n,n_0+\kappa_k}\,.
\end{align}
Here, $\kappa_k$ is a set of $N$ photon numbers with
$|\kappa_i - \kappa_j| > 1$ for all $1\leq i\neq j\leq N$.
Thus if for arbitrary $n$ one has $C_3^n(0)\neq 0$, 
then $C_3^{n+1}(0)=0$ and $C_3^{n-1}(0)=0$.
In this case, the result is equal to the
result for a single Fock mode, i.e. there is no 
interference independence of the width 
$[\max_k (\kappa_k) - \min_k (\kappa_k)]$ 
of the initial photon number
distribution.

\section{\label{sec-discussion}Discussion}
\begin{figure}[t!]
\includegraphics[width=8cm]{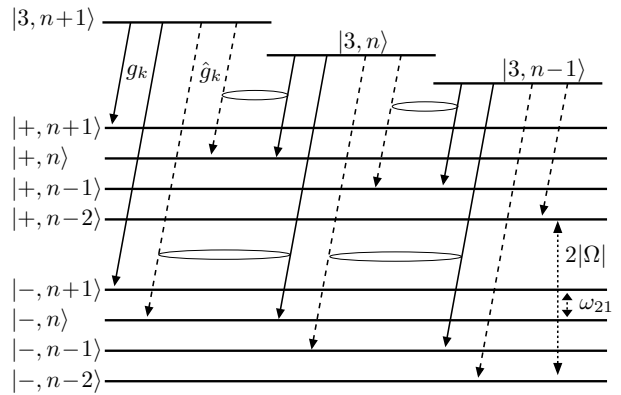}
\caption{\label{fig-decay}Decay pathways starting from single
upper state Fock modes $|3,n\rangle$ into the corresponding 
lower dressed states. The solid (dashed) arrows are decays due 
to coupling via coupling constant $g_k$ ($\hat{g}_k$). The ellipses
mark interfering pathways. The parameters are chosen as in 
Fig.~\ref{fig-reference}, such that the AC-Stark splitting 
$|\Omega|$ is larger than the photon frequency splitting 
$\omega_{21}$, as indicated by the double arrows.}
\end{figure}

In order to understand the results of the above 
Section~\ref{sec-numerics}, it is important to note
that even though the driving field is assumed to couple the
two lower states only, the upper state atomic state $|3\rangle$
also splits up in a multiplet of states $|3,n\rangle$ which 
decays into the corresponding dressed states.
Fig.~\ref{fig-decay} shows all the possible decay pathways from
three adjacent upper states $|3,n+1\rangle$, $|3,n\rangle$ and 
$|3,n-1\rangle$.
First, we assume that only state $|3,n\rangle$ is populated,
i.e. a single Fock mode. Then Fig.~\ref{fig-decay}
shows that each of the possible final states is only reached via
a single pathway from the initial state. Thus no interference is
possible, in accordance with the results in Section~\ref{sec-numerics}.
Next we assume that all three adjacent upper states $|3,n\rangle$ and
$|3,n\pm 1\rangle$ are populated. Then some of the final states
can be reached via two pathways, as indicated by the ellipses
in Fig.~\ref{fig-decay}, which gives rise to interference
effects. 
It is interesting to note that the two initial states
of the interfering pathways  are different, only the final state
is the same. The two initial states, however, have a fixed phase
relation with relative phase $\phi$, which nevertheless allows
for interference to take place. The interfering subsystems are thus
independent three-level systems in $V$ configuration. Note that
independent interference to several final states was also
found in other systems~\cite{EvBuKe2002}.
This conclusion can be further verified by noting from 
Fig.~\ref{fig-decay} that for each final state, one of the pathways 
is mediated by the coupling constant $g_k$, whereas the other path 
is proportional to $\hat{g}_k$. Thus the interference effects 
should also be controllable via the relative phase of these coupling 
constants just as via the phase $\phi$. A numerical check shows that
this is indeed the case.

\begin{figure}[t!]
\includegraphics[width=8cm]{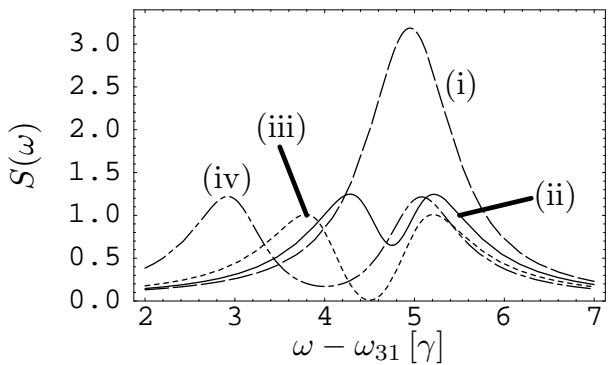}
\caption{\label{fig-w21-dep}Dependence of the interference effects
on the lower state separation $\omega_{21}$. The parameters are
as in Fig.~\ref{fig-reference}(a), except for (i) $\omega_{21} = 0.1$,
(ii) $\omega_{21} = 0.5$, (iii) $\omega_{21} = 1$,
(iv) $\omega_{21} = 2$.}
\end{figure}

The interference mechanism in the subsystems in $V$ configuration, however,
is not completely analogous to the well-known spontaneous-emission 
interference in $V$ systems with parallel transition dipole 
moments (see e.g.~\cite{ZhChLe1995}).
This is illustrated in Fig.~\ref{fig-w21-dep}. There, the 
interference structure around $\omega-\omega_{31} = 5$ in 
Fig.~\ref{fig-reference}(a) is shown again, but for different values
of $\omega_{21}$. It can be seen that perfect interference, i.e. a vanishing
of the spectral intensity at a certain frequency, only occurs for 
curve (iii) with
$\omega_{21}=\gamma$. The interference effects are reduced both
if $\omega_{21}$ is increased or decreased. This is in contrast to the
usual $V$ system with parallel dipole moment, where the interference 
effects only become smaller if the level spacing is increased. This 
upper restriction to $\omega_{21}$ warrants that the interfering pathways 
cannot be distinguished by their transition frequencies. 
In our  system, the additional requirement $\omega_{21}\approx \gamma$ for 
maximum interference stems from the fact that for total destructive
interference, the two amplitudes must have equal weights. This is the case 
for $\omega_{21}=\gamma$.
Thus the analogy to spontaneous decay in atomic $V$ systems with
non-orthogonal transition dipole moments in~\cite{ZhChLe1995}
is not complete.

In Section~\ref{sec-numerics}, we found that for small numbers
of adjacent Fock states initially populated, there is a quantitative
deviation of the spectrum from the classical case. The reason for this
is that the decay pathways which do not take part in an interfering 
$V$ subsystem disturb the spectrum. For a narrow distribution with 
equal weights, the relative weight of the disturbing pathways is not
negligible as compared to the weight of the interfering pathways such
that the quantum spectrum deviates from the classical one. If the number
of adjacent states initially populated grows, the relative weights of
the disturbing paths become small, and the spectrum approaches the 
classical one. A nice check is possible by considering 
three adjacent states $n, n\pm 1$ populated initially as in
Fig.~\ref{fig-decay}, but evaluating the spectrum with amplitudes 
$\alpha_n$ and $\beta_n$ only. The result is a spectrum identical
to the classical one up to a normalization factor $1/|C_3^n(0)|^2$,
as then only the interfering pathways are taken into account, while
the contribution of the disturbing final states which are only populated 
by a single pathway are neglected.

From Fig.~\ref{fig-decay} and from the definition of the dressed states
in Eq.~(\ref{dressedstates}) one may also understand why it is not 
possible to cancel the spontaneous emission completely.
On evaluating the transition matrix elements between upper states
$|3,n\rangle$ and the dressed states $|\pm,n\rangle$ for spontaneous
emissions, one finds that the relative phase between paths leading to
dressed states $|+,n\rangle$ differs by $\pi$ from the relative phase
between paths leading to dressed states $|-,n\rangle$. Thus
destructive interference is not possible to both dressed state manifolds
at the same time, and there is always spontaneous emission.

Finally, we discuss the case of distant Fock modes initially
populated. As can be seen from Fig.~\ref{fig-decay}, then none
of the final states can be reached via more than one pathway,
and no interference is possible. Thus it is not a large total width of
the initial driving field state, but rather the fixed phase
relation between adjacent Fock mode states, which accounts for the
interference effects.

In summary, we have discussed the spontaneous-emission spectrum
of a three-level system in $\Lambda$ configuration, which is sensitive 
to the phase of a quantized electromagnetic field coupling the two
lower atomic states. The quantum treatment together with a specific
choice of non-classical initial states for the driving field allows
to identify the mechanism leading to interference and the phase dependence.
We found that the quantized driving field creates an upper-state multiplet
of Fock number states. The system shows interference signatures if 
the initial state of the driving field contains adjacent Fock
number states with a fixed phase relation. The reason for this
condition is that then the full level spectrum contains
three-level subsystems in $V$ configuration which allow for pathways 
interference on the two transitions to the lower state. Each of these
$V$ systems is made up from two states of the upper state multiplet and
a lower dressed state.


\appendix

\section{\label{app-derivation}Derivation of Eq.~(\ref{sol-c3})}
In this appendix, we briefly outline the derivation of the
solution to the upper state amplitude equation of motion given
in Eq.~(\ref{sol-c3}).
For this, we introduce the variables
\begin{align}
X_k^n =& \frac{1}{\sqrt{2}}\,\left ( \alpha_k^n + \beta_k^n \right )\,,
\\
Y_k^n =& \frac{1}{\sqrt{2}}\,\left ( \alpha_k^n - \beta_k^n \right )\,.
\end{align}
Then, Eqs.~(\ref{eom-c})-(\ref{eom-beta}) can be written as
\begin{align}
\frac{d}{dt}\,C_3^n =&\sum_k g_k\,e^{-i \delta_k t}\,X_k^n 
+ i\,\sum_k\hat{g}_k \,e^{i\phi}\,e^{-i\hat{\delta}_k t}\,Y_k^n\,, 
\label{eq-c}\\
\frac{d}{dt}\,X_k^n =& -g_k^*\,e^{i \delta_k t}\, C_3^n - i\,|\bar{g}|\,
\sqrt{n}\,Y_k^{n-1}\,, \label{eq-x}\\
\frac{d}{dt}\,Y_k^n =& i\, \hat{g}_k^* \, e^{-i\phi}\,
e^{i \hat{\delta}_k t}\, C_3^n - i\,|\bar{g}|\,
\sqrt{n+1}\,X_k^{n+1}\,.\label{eq-y}
\end{align}
The equations for $X_k^n$ and $Y_k^n$ can be decoupled e.g. by applying
a derivation with respect to time to Eq.~(\ref{eq-x}) and inserting 
Eq.~(\ref{eq-y}), which yields:
\begin{align}
\frac{d^2}{dt^2}\,X_k^n =& -|g|^2\,n\,X_k^n - g_k^*\,\frac{d}{dt}\left (
e^{i\delta_k t}\,C_3^n\right ) \nonumber \\
&+ \hat{g}_k^*\,\bar{g}^*\,\sqrt{n}\,
e^{i\hat{\delta}_k t}\,C_3^{n-1} \label{eq-xx}\,.
\end{align}
The further treatment of this equation is simplified by noting that
$X_k^n(0)=0=Y_k^n(0)$, as we assume the atom to be in the excited state
$|3\rangle$ initially. 
We can then write the solution of Eq.~(\ref{eq-xx}) as
\begin{align}
X_k^n(t) &= -g_k^*\,\int_0^t \,e^{i\delta_k \tau}\,\cos[v_n(t-\tau)]\,
C_3^n(\tau)\,d\tau \nonumber \\
&+ \hat{g}_k^*\, e^{-i\phi}
\int_0^t \,e^{i\hat{\delta}_k \tau}\,\sin[v_n(t-\tau)]  \,
C_3^{n-1}(\tau)\,d\tau       \,,\label{sol-xx}
\end{align}
with $v_n = |g|\sqrt{n}$.
An analogous calculation yields
\begin{align}
&Y_k^n(t) = i\hat{g}_k^*\,e^{-i\phi}\,\int_0^t \,e^{i\hat{\delta}_k t}
\,\cos[v_{n+1}(t-\tau)]\,
C_3^n(\tau)\,d\tau \nonumber \\
& + i g_k^*\,e^{-i\phi} 
\,\int_0^t \,e^{i\delta_k t}
\,\sin[v_{n+1}(t-\tau)]\,
C_3^{n+1}(\tau)\,d\tau
\,. \label{sol-yy}
\end{align}
Inserting Eqs.~(\ref{sol-xx}), (\ref{sol-yy}) in Eq.~(\ref{eq-c}),
one obtains
\begin{align}
&\frac{d}{dt}\,C_3^n(t) = \nonumber \\
& -\sum_k |g_k|^2\, \int_0^t e^{-i\delta_k (t-\tau)}
  \cos[v_n(t-\tau)]\, C_3^n(\tau)\,d\tau \nonumber 
 \displaybreak[0]\\
& -\sum_k |\hat{g}_k|^2\, \int_0^t e^{-i\hat{\delta}_k (t-\tau)}\,
\cos[v_{n+1}(t-\tau)]\, C_3^n(\tau)\,d\tau\nonumber 
\displaybreak[0] \\
& + \sum_k  g_k\,\hat{g}_k^*\,e^{-i\phi}\,
\int_0^t e^{-i\delta_k (t-\tau)} \, e^{i\omega_{21}\tau}\nonumber \\
& \qquad \times \sin[v_{n}(t-\tau)]\, C_3^{n-1}(\tau)\,d\tau\nonumber 
\displaybreak[0] \\  
& - \sum_k  g_k^*\,\hat{g}_k\,e^{i\phi}\,
\int_0^t e^{-i\delta_k (t-\tau)}\,e^{-i\omega_{21}\tau} \nonumber \\
& \qquad \times \sin[v_{n+1}(t-\tau)]\, C_3^{n+1}(\tau)\,d\tau \,.
\label{upper-state}
\end{align}
Here, the last two terms are interference cross terms which are present
as the two transition dipole moments are assumed to be non-orthogonal.
Equation~(\ref{upper-state}) can be evaluated in a Wigner-Weisskopf-like 
analysis to
\begin{align}
\frac{d}{dt}\,C_3^n(t) = -\frac{\gamma}{2}\,C_3^n(t)\,,
\end{align}
where $\gamma = 2\pi [D(\omega_{31})\,g(\omega_{31})^2 
+ D(\omega_{32})\,\hat{g}(\omega_{32})^2]$ is the total upper state
decay rate.
Note that the Wigner-Weisskopf procedure gives rise to a delta function
$\delta(t-\tau)$ in the integrand, such that the interference cross 
terms containing a sine function in Eq.~(\ref{upper-state}) vanish.
Thus the upper state decay is not affected by the fact that the two
transition dipole moments are assumed to be parallel.

\end{document}